\documentclass[twocolumn,longbibliography,nofootinbib,aps,prl,superscriptaddress]{revtex4-2}    
\usepackage{CJK}
\pdfoutput=1 
\usepackage{soul}
\usepackage{mathtools,amssymb,amsfonts,amsthm,physics,subcaption,graphicx,multirow,float,bm}
\usepackage[colorlinks,citecolor=blue,linkcolor=blue,urlcolor=blue, breaklinks=true]{hyperref}
\usepackage[capitalize]{cleveref}
\usepackage[shortlabels]{enumitem}
\setlist{  
  listparindent=\parindent,
  parsep=0pt,
}
\counterwithout{figure}{section}
\usepackage[none]{hyphenat}
\usepackage{csquotes}

\usepackage{color}

\newcommand{\blk}{\color{black}}

\usepackage{bm}
\usepackage{dsfont}
\setlist{nosep}
\newcommand{\cfum}{Centro de F\'{i}sica, Universidade do Minho, Campus de Gualtar, 4710-057 Braga, Portugal} 
\newcommand{\inl}{INL -- International Iberian Nanotechnology Laboratory, Av. Mestre Jos\'{e} Veiga s/n, 4715-330 Braga, Portugal} 

\usepackage{thmtools} 
\usepackage{thm-restate}

\newcommand{\nocontentsline}[3]{}
\let\oldaddcontentsline\addcontentsline
\newcommand{\tocless}[2]{%
  \let\addcontentsline=\nocontentsline#1{#2}
  \let\addcontentsline\oldaddcontentsline}
  
\begin{document}


\title{Nonnegativity of a Kirkwood-Dirac representation implies noncontextuality}






\title{Kirkwood-Dirac representations beyond quantum states \\ (and their relation to noncontextuality)}





\author{David Schmid} 
 \email{davidschmid10@gmail.com}
\author{Roberto D. Baldijão}
\affiliation{International Centre for Theory of Quantum Technologies, University of Gda\'nsk, 80-308 Gda\'nsk, Poland}
\author{Y{\`i}l{\`e} Y{\=\i}ng}
\affiliation{Perimeter Institute for Theoretical Physics, 31 Caroline Street North, Waterloo, Ontario Canada N2L 2Y5}
\affiliation{Department of Physics and Astronomy, University of Waterloo, Waterloo, Ontario, Canada, N2L 3G1}
\author{Rafael Wagner}
\email{rafael.wagner@inl.int}
\affiliation{\inl}
\affiliation{\cfum}
\author{John H. Selby}
\affiliation{International Centre for Theory of Quantum Technologies, University of Gda\'nsk, 80-308 Gda\'nsk, Poland}

\begin{abstract}
Kirkwood-Dirac representations of quantum states are increasingly finding use in many areas within quantum theory. Usually, representations of this sort are only applied to provide a representation of quantum states (as complex functions over some set). We show how standard Kirkwood-Dirac representations can be extended to a fully compositional representation of {\em all} of quantum theory (including channels, measurements and so on), and prove that this extension satisfies the essential features of functoriality (namely, that the representation commutes with composition of channels), linearity, and quasistochasticity. Interestingly, the representation of a POVM element is uniquely picked out to be the collection of weak values for it relative to the bases defining the representation. We then prove that if one can find any Kirkwood-Dirac representation that is everywhere real and nonnegative for a given experimental scenario or fragment of quantum theory, then the scenario or fragment is consistent with the principle of generalized noncontextuality, a key notion of classicality in quantum foundations. 
We also show that the converse does {\em not} hold: even if one verifies that all Kirkwood-Dirac representations (as defined herein) of an experiment require negativity or imaginarity, one cannot generally conclude that the experiment witnesses contextuality.
\end{abstract}

\maketitle

\raggedbottom

\section{Introduction} 


It is often useful to represent quantum systems as functions over some kind of phase space.
If one wishes to represent all of standard quantum theory, however, well-known no-go theorems demonstrate that such representations must be \emph{quasi}probabilistic rather than probabilistic~\cite{Ferrie_2008,negativity}. Often, the term `\emph{quasi}' refers to the fact that these functions take negative values, with well-known examples including Wigner quasiprobability distributions~\cite{wigner1932quantum} or discrete versions thereof~\cite{gross2006hudson}. 
However, one intriguing family of quasiprobability distributions of quantum states that is defined over the {\em complex} numbers is the Kirkwood-Dirac (KD) quasiprobability distribution~\cite{kirkwood1933quantum,dirac1945analogy,arvidssonshukur2024properties}. KD distributions have been used to study post-selected quantum metrology~\cite{arvidsson_Shukur2020quantum,lupu_Gladstein2022negative,jenne2022unbounded}, quantum  fluctuation theorems~\cite{lostaglio2018fluctuation,halpern2017jarzynski,upadhyaya2023happens}, work extraction~\cite{baumer2018fluctuating,lostaglio2022kirkwood,hernandezgomez2023projective,gherardini2024quasiprobabilities}, heat flow~\cite{lostaglio2020quasiprobability}, weak value theory and quantum coherence estimation~\cite{wagner2023simple,budiyono2023quantifying}, indefinite causal order~\cite{ban2021sequential}, incompatibility~\cite{gao2023measuring,arvidsson_Shukur2021conditions,deBievre2021complete}, and scrambling of quantum information~\cite{halpern2018quasiprobability,gonzalez2019otoc,alonso2022diagnosing}. Moreover, KD distributions can be experimentally measured in various ways~\cite{wagner2024circuits,lostaglio2022kirkwood}. 
Perhaps surprisingly, given the recent surge of interest in such representations, KD distributions have so far only been described for quantum states, in contrast to the  majority of other established representations (such as Wigner's) that can be used to represent quantum dynamics, quantum measurements, and indeed arbitrary quantum processes. 

The first contribution of this work is to extend KD distributions as typically defined---for quantum states only---to full \emph{representations of quantum theory}, applicable to arbitrary quantum processes---states, channels, measurements, instruments, and so on.
Our extension is designed to satisfy a powerful and elegant mathematical property: that the representation commutes with composition. Formally, the representation is a symmetric strict monoidal functor~\cite{mac2013categories} (i.e., is diagram-preserving~\cite{Schmid2024structuretheorem}).
Our arguments here are {\em the} natural extension of work regarding real-valued quasiprobability representations, most notably Refs.~\cite{Ferrie_2008,Ferrie_2009,negativity,posquasi,Schmid2024structuretheorem}, to the {\em complex}-valued family of KD representations. Indeed, all of these constructions are special cases of what are known as {\em frame representations}~\cite{christensen2016introduction}, and this fact immediately unifies much of the study of KD representations and other representations (such as the Wigner function). 

Our representation may be of independent interest to research on weak values~\cite{aharonov1988weak,tamir2013intro,dressel2014colloquium}. In particular, we show that the representation of a POVM element in a particular KD representation is given by the collection of weak values for that operator with respect to the basis vectors defining the KD representation. This implies a formal duality between KD representations for states and weak values.

It is sometimes suggested that when a quantum state's KD distribution has negative real components or non-zero imaginary components, respectively termed \emph{negativity} and \emph{imaginarity}, then that state is somehow nonclassical. 
But (as has been argued before~\cite{negativity}) it is meaningless to talk about the nonclassicality of a state without specifying more information---e.g., the measurements one can do on the state or the experiment one will embed the state in. By considering the representation of arbitrary processes (not merely quantum states), we begin to derive direct connections between KD representations and an important foundational notion of classicality, namely generalized noncontextuality. Generalized noncontextuality~\cite{gencontext} (henceforth simply `noncontextuality') is a principled~\cite{gencontext,SchmidGPT,Leibniz}, useful~\cite{schmid2018contextual,AWV,KLP19,POM,RAC,RAC2,Saha_2019,cloningcontext,comp1,comp2,wagner2024coherence, Yadavalli2020,saha2019preparation,contextmetrology}, and operational~\cite{operationalks,mazurek2016experimental,giordani2023experimental, Kunjwal16,Schmid2018,pusey2019contextuality,selby2022open,selby2023accessible,selby2024linear,schmid2024addressing} notion of classicality, relative to which one can classify {\em any} theory or experiment as classically-explainable or not. Indeed, this classification has been studied for phenomena ranging from computation~\cite{Schmid2022Stabilizer,shahandeh2021quantum}, state discrimination~\cite{schmid2018contextual,flatt2021contextual,mukherjee2021discriminating,Shin2021}, interference~\cite{Catani2023whyinterference,catani2022reply,catani2023aspects}, compatibility~\cite{selby2023incompatibility,selby2023accessible}, uncertainty relations~\cite{catani2022nonclassical}, metrology~\cite{contextmetrology}, thermodynamics~\cite{contextmetrology}, weak values~\cite{AWV, KLP19}, coherence~\cite{rossi2023contextuality,wagner2024inequalities}, quantum Darwinism~\cite{baldijao2021noncontextuality}, information processing and communication~\cite{POM,RAC,RAC2,Saha_2019,Yadavalli2020}, cloning~\cite{cloningcontext}, and (as mentioned above) Bell~\cite{Wright2023invertible,schmid2020unscrambling} and Kochen-Specker correlations~\cite{operationalks,kunjwal2018from,Kunjwal16,Kunjwal19,Kunjwal20,specker,Gonda2018almostquantum}.

The second main contribution of this article is to prove that if an experiment or theory admits of {\em any} Kirkwood-Dirac representation that is everywhere real and nonnegative, then that theory or experiment is consistent with noncontextuality. 
We also prove that the converse does not hold---so that even if {\em every} KD representation (as defined herein) requires negativity or imaginarity, one has not necessarily proven the failure of noncontextuality.

\section{Kirkwood-Dirac representations of states, channels, and measurements}

Given a finite dimensional Hilbert space $\mathcal{H}_A$ associated to a single quantum system $A$, pick any two bases $\{\ket{a_i}\}_i$ and $\{\ket{a'_{i'}} \}_{i'}$ for the Hilbert space, under the constraint that $\braket{a_i}{a'_{i'}}\neq 0$ for all $i$ and $i'$. From these, define a (nonovercomplete) basis 
$F\coloneq\{F_{i,i'}\}_{i,i'}$ for the space of bounded linear operators $\mathcal{B}(\mathcal{H}_A)$ as \begin{align} 
\label{framedefn}
  F_{i,i'} \coloneq \ketbra{a'_{i'}}{a_i} \braket{a'_{i'}}{a_i}.
\end{align}
Such a basis is an example of a  {\em frame}~\cite{christensen2016introduction}. 
Next, define the \emph{dual frame} $D\coloneq\{D_{i,i'}\}_{i,i'}$ as the unique basis of operators satisfying 
\begin{align}
\label{eq:DualityFD}
    \Tr \left[F_{\bar{i},\bar{i}'}D_{i,i'}\right]=\delta_{\bar{i},i}\delta_{\bar{i}',i'},
\end{align}
namely,
\begin{align} 
\label{dualdefn}
  D_{i,i'} \coloneq \frac{\ketbra{a_i}{a'_{i'}}}{\braket{a'_{i'}}{a_i}}.
\end{align}
The frame and the dual frame operators satisfy
\begin{align}
\label{eq:TraceD}
&\Tr \bigl[D_{i,i'}\bigr] = 1,\\
&\Tr \left[F_{i,i'}\right] = |\!\braket{a'_{i'}}{a_i}\!|^2.
\end{align}

Using these, we construct the representations of state $\rho$ and POVM element $E$, respectively, as
\begin{align}
\label{eq:muFromrho}
  \mu(i,i'|\rho) \coloneq \Tr [ F_{i,i'} \rho]
\end{align}
and
\begin{align}
\label{eq:xiFromE}
  \xi(E|i,i') \coloneq \Tr [ E D_{i,i'}].
\end{align}
Given a channel $\mathcal{E}: \mathcal{B}(\mathcal{H}_A) \rightarrow \mathcal{B}(\mathcal{H}_B)$, one defines a frame $F_{j,j'}$ and its dual $D_{j,j'}$ for $\mathcal{H}_B$ in the same manner as above (but relative to two bases for $\mathcal{H}_B$, denoted $\{\ket{b_j}\}_j$ and $\{\ket{b'_{j'}} \}_{j'}$, satisfying $\braket{b_j}{b'_{j'}}\neq 0,\forall j,j'$), in addition to those introduced above for system $\mathcal{H}_A$, with which the channel is represented by 
\begin{align}
\label{eq:GammaFromChannel}
  \Gamma(j,j'|i,i',\mathcal{E}) \coloneq &\Tr [F_{j,j'} {\cal E}( D_{i,i'})]. 
  \end{align}
It is often useful to interpret $\Gamma(j,j'|i,i',\mathcal{E})$ as a matrix mapping vectors indexed by $(i,i')$ to vectors indexed by $(j,j')$.
Composition within the representation is then given by matrix multiplication.
The family of all KD representations is generated by ranging over possible basis choices for each Hilbert space in question.

Note that while we use the suggestive notation typically used in the literature on ontological models for epistemic states, response functions, and stochastic maps, these are more general, as they are complex-valued and not bounded by 0 or 1.

Such representations are faithful (injective).\footnote{One could consider representations that are not faithful, but such representations generally lose information; moreover, faithfulness is {\em necessary} for reproducing the experimental predictions arising in scenarios where the states and measurement operators span the space $\mathcal{B}(\mathcal{H})$.} That is, given some representation $\mu$, $\xi$, or $\Gamma$, one can reconstruct the quantum processes $\rho$, $E$ or $\mathcal{E}(\cdot)$, respectively, by 
\begin{align}
\label{eq_recrho}
  \rho = & \sum_{i,i'} \mu(i,i'|\rho) D_{i,i'}; \\
  \label{eq_recE}
  E = & \sum_{i,i'} \xi(E|i,i')  F_{i,i'};  \\
\label{eq_recChann}
    {\cal E}(\cdot)= & \!\!\!\sum_{j,j',i,i'} \!\!\! \Gamma(j,j'|i,i',\mathcal{E}) \Tr [F_{i,i'} \ \cdot\ ] D_{j,j'},
\end{align}
as we prove in Appendix~\ref{subsec:MathStuff}.

To see that the quantum predictions are recovered by the representation, consider an experiment where a channel $\mathcal{E}$ is carried out on state $\rho$, after which a measurement outcome corresponding to POVM element $E$ is obtained. The representation of this scenario is
\begin{align}
&\sum_{j,j',i,i'} \xi(E|j,j') \Gamma(j,j'|i,i',\mathcal{E}) \mu(i,i'|\rho) \nonumber \\
  = & \sum_{j,j',i,i'} \xi(E|j,j') \Tr [F_{j,j'} {\cal E}( D_{i,i'})] \mu(i,i'|\rho) \nonumber \\
  = & \Tr \left[ \left(\sum_{j,j'}  \xi(E|j,j') F_{j,j'}\right)  {\cal E} \left( \sum_{i,i'} \mu(i,i'|\rho)D_{i,i'}\right) \right]\nonumber\\
  = & \Tr (E {\cal E}(\rho)),
\end{align}
where we used the linearity of the trace operation and of the channel for the second equality, and used \cref{eq_recrho}  and \cref{eq_recE} for the last equality.

The representation of the identity channel $\mathcal{I}$ is given by Eq.~\eqref{eq:GammaFromChannel} to be the identity matrix:
\begin{align} \label{repnidentity}
&\Gamma(\bar{i},\bar{i}'|i,i',\mathcal{I}) = \Tr \left[F_{\bar{i},\bar{i'}}\mathcal{I}(D_{i,i'})\right] \nonumber\\ 
&= \Tr \left[F_{\bar{i},\bar{i}'}D_{i,i'}\right] = \delta_{\bar{i},i}\delta_{\bar{i}',i'},
\end{align}
as per Eq.~\eqref{eq:DualityFD}.
Finally, the representation of the identity POVM element is given by 
Eq.~\eqref{eq:xiFromE} to be the all-ones vector: for all $i,i'$, 
\begin{align} 
\label{unitrepn}
    \xi(\mathds{1}|i,i') = \Tr \left[\mathds{1}D_{i,i'}\right] = \frac{\Tr  \left[\ketbra{a_i}{a'_{i'}} \right]}{\braket{a'_{i'}}{a_i}} = 1.
\end{align}
This condition is the reason it is meaningful to think of these representations as quasistochastic (i.e., where the sum of the components in each column of one's matrix is less than or equal to one), as opposed to some arbitrary complex linear representation, since it implies that the representation of every trace-preserving channel $\mathcal{E}$ satisfies
\begin{align}\label{tpnormcond0}
\sum_{j,j'} \Gamma(j,j'|i,i',\mathcal{E})
= 1 \quad \quad \forall i,i' 
\end{align}
and, for the special case of a state $\rho$, satisfies
\begin{equation}\label{statenorm0}
\sum_{i,i'}\mu(i,i'|\rho) = 1.
\end{equation}
The proof is simple, and given in Appendix~
\ref{subsec:MathStuff}. 

Our choice of normalisation factors for the frame in Eq.~\eqref{framedefn} might have seemed arbitrary, but in fact it is fixed, since (by Eq.~\eqref{eq:DualityFD}) our choice fixes the normalization for the dual, which in turn is fixed by Eq.~\eqref{unitrepn}. This is one reason why it must be the case that $\braket{a'_{i'}}{a_i}\neq 0$ for all $i$ and $i'$. (See  Appendix~\ref{subsec:MathStuff} for more.) Note in particular that this rules out taking the bases $\{\ket{a_i}\}_i$ and $\{\ket{a'_{i'}} \}_{i'}$ to be the same basis, which would lead to the standard density matrix representation of quantum states (for which Eq.~\eqref{unitrepn} is not satisfied).  

\subsection{Multiple systems}

For composite systems, we take the operator basis to be the tensor product of the bases of the space of operators on each Hilbert space in the composite. For example, for the composite Hilbert space $\mathcal{H}_A \otimes \mathcal{H}_B$, we pick any two bases $\{\ket{a_i}\}_i$ and $\{\ket{a'_{i'}} \}_{i'}$ satisfying $\braket{a_i}{a'_{i'}}\neq 0, \forall i, i'$ for $\mathcal{H}_A$ and any two bases $\{\ket{b_j}\}_j$ and $\{\ket{b'_{j'}} \}_{j'}$ satisfying $\braket{b_j}{b'_{b'}}\neq 0, \forall j, j'$ for $\mathcal{H}_B$. We   introduce the frame 
\begin{align}
\label{eq_def2frame}
 F_{i,i';j,j'} \coloneq 
  & \ketbra{a'_{i'}}{a_i}\otimes\ketbra{b'_{j'}}{b_j} \braket{a'_{i'}}{a_i}\braket{b'_{j'}}{b_j}  \\
  = & F_{i,i'}\otimes F_{j,j'} \nonumber
\end{align}
and the dual frame 
\begin{align}
\label{eq_def2dual}
    D_{i,i';j,j'} \coloneq 
    &\frac{ \ketbra{a_i}{a'_{i'}} \otimes \ketbra{b_j}{b'_{j'}} } { \braket{a'_{i'}}{a_i} \langle{b'_{j'}}\vert{b_j}\rangle }  \\
    =& D_{i,i'}\otimes D_{j,j'} \nonumber
\end{align}
and define the representation of quantum processes exactly as was done for single systems.

\subsection{Recovering standard Kirkwood-Dirac distributions for states}

For the special case of quantum states, our representation recovers the standard KD distribution~\cite{wagner2024circuits,arvidssonshukur2024properties,halpern2018quasiprobability}, since
\begin{align*}
  \mu(i,i'| \rho) \coloneq & \Tr [ F_{i,i'} \rho]= \bra{a_i}\rho\ket{a'_{i'}}\braket{a'_{i'}}{a_i}.
\end{align*}
Consequently, our representation satisfies all known properties of preexisting KD representations; for instance, the distribution associated to any state is bounded by one and has maximal negativity and imaginarity given by $\min_{\{a_i\}_i,\{a'_{i'}\}_{i'},\rho} \text{Re}[\mu(i,i'|\rho)] = -1/8 $ and $\max_{\{a_i\}_i,\{a'_{i'}\}_{i'},\rho} \text{Im}[\mu(i,i'|\rho)] = 1/4$~\cite{fernandes2024unitaryinvariant}; see Appendix~\ref{subsec:MathStuff} for more. Consequently, existing techniques to witness KD negativity or imaginarity, or that investigate the geometry of nonnegatively represented states with respect to fixed KD bases~\cite{arvidsson_Shukur2021conditions,deBievre2021complete,deBievre2023relating,xu2024kdclassical,wagner2024circuits,langrenez2023characterizing} also remain applicable. Our work opens the possibility of extending these investigations beyond quantum states.

 \subsection{Connections with weak values and tomography} 

While $\mu(i,i'|\rho)$ is bounded, the representation $\xi(E|i,i')$ is not, possibly having arbitrarily large positive, negative, or imaginary values. This is immediate from noticing that, for any dual frame $D$ given by Eq.~\eqref{dualdefn},  each value $\xi(E|i,i')$ is exactly what is known as the weak value of $E$ with respect to the vectors $\vert a_i\rangle$ and $\vert a'_{i'}\rangle$~\cite{dressel2014colloquium,tamir2013intro}.  Indeed, KD distributions for states and these particular weak values \emph{must} be dual representations of states and POVM elements in our representation, as the dual basis is uniquely singled out by Eq.~\eqref{eq:DualityFD}.  It remains to be seen how this duality relates to prior connections between weak values and KD distributions~\cite{lostaglio2022kirkwood,wagner2023simple,halpern2018quasiprobability}.

The KD representation $\Gamma$ of a channel $\mathcal{E}$, given by Eq.~\eqref{eq:GammaFromChannel}, can also have arbitrarily large absolute values due to the inner-products $\langle a'_{i'}|a_i\rangle$ in the denominator. Still, the quasistochastic behavior from  Eq.~\eqref{tpnormcond0} continues to hold. A natural question is how these representations relate to physical quantities of interest in the same way that those for states have been shown to, or as we have just argued that those for POVM elements relate to weak values.

As a final remark, since our representation is faithful and so contains complete information about one's quantum process, any estimation of our KD representation for a state, channel, or measurement is a form of state tomography, process tomography, or measurement tomography. For any process, the representation can be written as a  multivariate trace, and so can be estimated using polynomial~\cite{oszmaniec2024measuring,yosef2024multivariate} or even constant~\cite{quek2024multivariatetrace} depth quantum circuits.

\subsection{Compositionality and functoriality}

A useful property for any representation to have is functoriality, or more specifically, that the representation is a symmetric strict monoidal functor~\cite{mac2013categories} (sometimes called a diagram preserving map \cite{Schmid2024structuretheorem}),
which means that the representation commutes with composition of processes. Functoriality also requires that identity and swap operations are represented by the identity and the swap, respectively, in the representation.  Functoriality is arguably {\em essential} if that representation is to carry physical meaning (e.g. to provide a physical explanation) rather than merely be a mathematical simulation. Most representations studied to date are functorial, although some~\cite{husimi1940distribution,glauber1963coherent,sudarshan1963equivalence} are not~\cite{Schmid2024structuretheorem}.

As we have defined it, every KD representation is functorial. We prove this in Appendix~\ref{DPproof}, by showing that the representation of a composite process (whether composed in parallel or in sequence) is the same as the composite of the representations of each component of it, and by proving that the identity and swap operations are represented correctly.
For example, the representation of two channels composed in sequence, e.g. $\mathcal{E}_2\circ\mathcal{E}_1$, must be the composition (matrix product) of the representations of each individually, namely
\begin{align}
    & \Gamma(k,k'|i,i',\mathcal{E}_2\circ\mathcal{E}_1) \nonumber \\
    = & \sum_{j,j'}  \Gamma(k,k'|j,j',\mathcal{E}_1)\Gamma(j,j'|i,i',\mathcal{E}_2). \label{eq_Gammamu}
\end{align}

\section{Connection to generalized noncontextuality}


With Kirkwood-Dirac representations for all of quantum theory in hand, it is straightforward to establish connections with noncontextuality. 

Consider any fragment of quantum theory. In the simplest case, this means a set $\{\rho_i\}_i$ of states, a set $\{\mathcal{E}_j\}_j$ of channels, and a set $\{\{E_{[k|m]}\}_k \}_m$ of measurements. More generally, one may have any fixed circuit and a set of possibilities for each process in the circuit. These sets may arise from the particular capabilities of some particular hardware, from a specific experiment, or from some theoretical considerations. They may (but need not) have further structure, such as being finite or convexly closed. 

An {\em ontological model} for a fragment of quantum theory is any linear and functorial mapping from the quantum processes to (the monoidal category of) substochastic matrices \cite{Schmid2024structuretheorem}, such that the trace operation is mapped to marginalization (i.e., the all ones vector). See Ref.~\cite{Harrigan} for an introduction to ontological models, and Ref.~\cite{Schmid2024structuretheorem} for the first fully precise and compositional definition. Many fragments of quantum theory do not admit of any ontological model, but some fragments do---even very rich fragments like the stabilizer subtheory in all odd dimensions~\cite{gross2006hudson,Schmid2022Stabilizer}. 

Under the formulation we have introduced herein, any KD representation that happens to be real and nonnegative for all processes in a given fragment {\em also} satisfies all of these properties, and so constitutes an ontological model of that fragment. Clearly the KD representation in this case is a mapping from quantum processes to nonnegative real-valued matrices; as we prove in Appendix~\ref{stochproof}, these matrices are moreover {\em substochastic}.
And as we saw in Eq.~\eqref{repnidentity} and Eq.~\eqref{unitrepn} (respectively), the identity is represented by the identity matrix and the representation of the trace operation is equal to marginalization.

The connection to generalized noncontextuality now follows immediately from the results of Refs.~\cite{SchmidGPT,Schmid2024structuretheorem}, which show that the predictions of a fragment of quantum theory can be reproduced by an ontological model if and only if any laboratory procedures whose quantum description is given by this fragment admits of a generalized noncontextual ontological model.~\footnote{Formally, by `laboratory procedures' we mean processes within an unquotiented operational theory, while by the `quantum description' of the fragment, we mean a quotiented operational theory (with density operators, quantum channels, and the like).~\cite{chiribella2010probabilistic,Schmid2024structuretheorem}} In other words, a scenario is consistent with generalized noncontextuality if and only if its description within standard quantum theory admits of an ontological model. But every nonnegative KD representation constitutes an ontological model, so the existence of any nonnegative KD representation implies noncontextuality.

However, negativity or imaginarity of a KD representation does {\em not} necessarily imply the failure of noncontextuality. Most obviously, there may be another KD representation that manages to represent all processes in the scenario or theory nonnegatively, in which case this other representation constitutes a perfectly fine classical model, consistent with noncontextuality. 
Moreover, {\em even} if one shows that for a given scenario or theory, {\em all} KD representations (as defined herein) require negativity or imaginarity, it is {\em still} possible that an ontological model of an entirely different form is possible. 

Indeed, this sometimes happens, as we prove by example. Consider the stabilizer subtheory of quantum mechanics in odd dimensions, which has a {\em unique}~\cite{Schmid2022Stabilizer} ontological model, namely that given by Gross and by Spekkens~\cite{gross2006hudson,epistricted}. This model is not of the form of any Kirkwood-Dirac representation, as one can see by the fact that the frame operators defining it are Hermitian, while those of a KD representation are not.
But by uniqueness, there can be no other ontological model of it, so {\em all} KD representations of it must have negativity or imaginarity, even though the subtheory is noncontextual.

That said, there is a natural (although rather broad) generalization of the family of KD representations considered here and in the literature for which one can make a tighter connection with noncontextuality. This is the natural class of {\em linear, functorial, and empirically adequate quasistochastic} representations 
of quantum processes
as complex-valued functions: essentially, it is what one obtains if one relaxes the frame in Eq.~\eqref{framedefn} to be an arbitrary nonovercomplete frame (basis) for the space of all operators on $\mathcal{H}$. As we show in future work, one {\em can} prove the failure of noncontextuality by proving that every representation in this considerably broader class has negativity or imaginarity, as a natural extension of analogous arguments for real-valued functions in Ref.~\cite{Schmid2024structuretheorem}.

\section{Conclusions}

We have showed how to generalize standard KD representations to any quantum process, including measurements and channels. The representation satisfies key properties such as respecting composition and linearity. This allows us to clarify the relationship between KD distributions, ontological models, and noncontextuality. It also leads to new avenues of study, such as the connection of the KD representation of POVM elements with weak values.

Other extensions of KD distributions have been proposed, but only for quantum states~\cite{halpern2018quasiprobability,lupu_Gladstein2022negative,halpern2017jarzynski}. We are not aware of a fully general definition of a representation of quantum theory (or even of quantum states) in complex-valued functions. In forthcoming work, we extend the present work to define this very general class of representations and discuss their relation with some other generalized KD representations one can find in the literature. We then show that every ontological model of a quantum scenario has the form of some such representation that is nonnegative on all processes in the scenario, which implies that one can prove nonclassicality (the failure of generalized noncontextuality) if one can prove that no such representation is nonnegative. We also show that these general representations and results (which subsume all the ones in this paper) apply to arbitrary generalized probabilistic theories~\cite{Hardy,GPT_Barrett}, not just quantum theory. Finally, we note that working with representations such as these is typically dramatically simpler when one uses diagrammatic notation, as we do in the forthcoming work, and as is done in Ref.~\cite{Schmid2024structuretheorem}.

\vspace{3mm}

{\em Acknowledgements---} DS and JHS  were supported by the National Science Centre, Poland (Opus project, Categorical Foundations of the Non-Classicality of Nature, project no. 2021/41/B/ST2/03149). RDB acknowledges support by the Digital Horizon Europe
project FoQaCiA, Foundations of quantum computational advantage, GA No. 101070558, funded by the European Union. YY was supported by Perimeter Institute for Theoretical Physics. Research at Perimeter Institute is supported in part by the Government of Canada through the Department of Innovation, Science and Economic Development and by the Province of Ontario through the Ministry of Colleges and Universities. YY was also supported by the Natural Sciences and Engineering Research Council of Canada (Grant No. RGPIN-2024-04419). RW acknowledges support by FCT -- Fundação para a Ciência e a Tecnologia (Portugal) through PhD Grant SFRH/BD/151199/2021. 

\bibliography{reference.bib}

\appendix

\section{Useful mathematical facts}
\label{subsec:MathStuff}

We now present some useful mathematical facts regarding the family of Kirkwood-Dirac representations that we have defined. 

In the main text, we showed how arbitrary quantum processes could be represented and then reconstructed. In fact, arbitrary operators on a Hilbert space can be represented and reconstructed in an exactly analogous fashion. There are two ways to do so, depending on whether the representation or the reconstruction is done using the frame (as opposed to the dual). One can define the representation of an arbitrary operator $O$ acting on Hilbert space $\mathcal{H}_A$ as
\begin{align}
\label{repnO}
  \mu(i,i'|O) \coloneq & \Tr [ F_{i,i'} O],  \\
 = & \bra{a_i}O\ket{a'_{i'}}\braket{a'_{i'}}{a_i}, \label{eq:muexplicit}
\end{align}
where \cref{eq:muexplicit} has the form of standard KD representations commonly seen in the literature.
In this case, one reconstructs $O$ via
\begin{equation}
\label{eq_recOmu}
O =  \sum_{i,i'} \mu(i,i'|O) D_{i,i'}.
\end{equation}
The above was already noticed in, for example, Ref.~\cite[Eq. (19), pg. 4]{halpern2018quasiprobability}. Alternatively, one can represent $O$ as 
\begin{align}
\label{dualrepnO}
  \xi(O|i,i') \coloneq &\Tr [ O D_{i,i'}] \\
  = &  \frac{\bra{a'_{i'}}O\ket{a_i}}{\braket{a'_{i'}}{a_i}}; \label{eq:xiexplicit} 
\end{align}  
in which case the reconstruction equation is
\begin{equation}
\label{eq_recOxi}
  O = \sum_{i,i'} \xi(O|i,i')  F_{i,i'}.  
\end{equation}
Here we are using the convention that representations in terms of the frame are denoted by $\mu$, while those in terms of the dual frame are represented by $\xi$, consistent with the notation in the main text.

We now prove the validity of the reconstruction equations \cref{eq_recrho,eq_recE,eq_recChann} in the main text, of which the proof for  \cref{eq_recrho} and \cref{eq_recE} also works for the respective reconstruction equations for a general operator, namely, \cref{eq_recOmu} and \cref{eq_recOxi}.

For \cref{eq_recrho}, i.e., $  \rho = \sum_{i,i'} \mu(i,i'|\rho) D_{i,i'}$:
        \begin{align}
            & \sum_{i,i'} \mu(i,i'|\rho) D_{i,i'} \nonumber \\
            = & \sum_{i,i'} \bra{a_i}\rho\ket{a'_{i'}}\braket{a'_{i'}}{a_i} \frac{\ketbra{a_i}{a'_{i'}}}{\braket{a'_{i'}}{a_i}} \nonumber \\
            = & \sum_{i,i'}  \ketbra{a_i}{a_i}\rho\ketbra{a'_{i'}}{a'_{i'}} \nonumber \\
            = & \rho. \label{eq:proofrho}
        \end{align}
    When we replace $\rho$ with a general operator $O$, \cref{eq:proofrho} becomes a proof for \cref{eq_recOmu}.

For \cref{eq_recE}, i.e., $ E =  \sum_{i,i'} \xi(E|i,i')  F_{i,i'}$: 
        \begin{align}
            & \sum_{i,i'} \xi(E|i,i')  F_{i,i'} \nonumber \\
            = & \sum_{i,i'} \frac{\bra{a'_{i'}}E\ket{a_i}}{\braket{a'_{i'}}{a_i}} \ketbra{a'_{i'}}{a_i} \braket{a'_{i'}}{a_i} \nonumber \\
            = & \sum_{i,i'} \ketbra{a'_{i'}}{a'_{i'}}E\ketbra{a_i}{a_i} \nonumber \\
            = & E. \label{eq:proofE}
        \end{align}
    When we replace $E$ with a general operator $O$, \cref{eq:proofE} becomes a proof for \cref{eq_recOxi}.

 To prove \cref{eq_recChann}, namely that
\begin{equation}
\nonumber
    {{\cal E}(\cdot) = \!\!\!\!\! \sum_{j,j',i,i'}  \!\!\! \Gamma(j,j'|i,i',\mathcal{E})\Tr [F_{i,i'} \ \cdot \ ] \!D_{j,j'}},
\end{equation} 
we have
        \begin{align}
        & \!\!\! \sum_{j,j',i,i'} \!\!\! \Gamma(j,j'|i,i',\mathcal{E}) \Tr [F_{i,i'} \ \cdot \ ] D_{j,j'} \nonumber \\
        = &  \!\!\! \sum_{j,j',i,i'} \!\!\! \Tr [F_{j,j'} {\cal E}( D_{i,i'})] \mu(i,i'|\ \cdot\ ) D_{j,j'} \nonumber \\
        = & \sum_{j,j'} \Tr [F_{j,j'} {\cal E} \left(  \sum_{i,i'} \mu(i,i'|\ \cdot\ ) D_{i,i'}\right)] D_{j,j'} \nonumber \\
        = & \sum_{j,j'} \Tr [F_{j,j'} {\cal E} \left(  \cdot\right)] D_{j,j'} \nonumber \\
        = &   \sum_{j,j'}  \mu\left(j,j'|\mathcal{E} (\cdot)\right) D_{j,j'} \blk \nonumber \\
        = & {\cal E}(\cdot),
        \end{align}
        where we used the linearity of trace and channel at the second equal sign, and used the reconstruction equations at the third and fifth equal sign.
One relation that will be useful in the following proofs is that $\sum_{i,i'} F_{i,i'}$ preserves the trace of any operator. That is,
\begin{align}
\label{eq:SumFPreservesTrace}
    \sum_{i,i'}\Tr \left[F_{i,i'}O\right] = \Tr \left[O\right].
\end{align}
This can be seen by explicit calculation:
\begin{align}
    & \sum_{i,i'}\Tr \left[F_{i,i'}O\right] =\sum_{i,i'} \Tr \Bigl[\braket{a'_{i'}}{a_i}\ketbra{a'_{i'}}{a_i} O\Bigr] \nonumber\\
        = & \sum_{i,i'}\Tr \Bigl[\ket{a'_{i'}}\!\!\braket{a'_{i'}}{a_i}\!\!\bra{a_i} O\Bigr] \nonumber\\
        = & \Tr \Bigl[\Bigl(\sum_{i'}\ketbra{a'_{i'}}{a'_{i'}}\Bigr)\Bigl(\sum_{i}\ketbra{a_i}{a_i}\Bigr) O\Bigr] \nonumber \\
        = & \Tr \left[O\right].
\end{align}
In particular, this immediately implies that the KD representation of an arbitrary quantum state (or indeed of an arbitrary trace-$1$  operator when using the frame rather than the dual representation) satisfies the normalization property in Eq.~\eqref{statenorm0} of the main text,
\begin{equation}\label{statenorm}
\sum_{i,i'}\mu(i,i'|\rho) = 1.
\end{equation}

In addition, the representation of a trace-preserving channel in every KD representation satisfies Eq.~\eqref{tpnormcond0} of the main text, namely
\begin{align}\label{tpnormcond}
\sum_{j,j'} \Gamma(j,j'|i,i',\mathcal{E})
= 1 \quad \quad \forall i,i', \mathcal{E}
\end{align}
This follows from 
    \begin{align}
        &\sum_{j,j'} \Gamma(j,j'|i,i',\mathcal{E}) 
        =\sum_{j,j'}\Tr \left[F_{j,j'}\mathcal{E}(D_{i,i'})\right] \nonumber \\
        =& \Tr  \left[\mathcal{E}(D_{i,i'})\right]  
        =  \Tr  \left[\mathds{1}\mathcal{E}(D_{i,i'})\right]\nonumber\\
        =&\Tr  \left[\mathcal{E}^{\dagger}(\mathds{1})D_{i,i'}\right] = \Tr  \left[D_{i,i'}\right] = 1,
    \end{align}
where in the second equality we used Eq.~\eqref{eq:SumFPreservesTrace}, in the penultimate step we used that $\mathcal{E}$ is trace-preserving, which implies its adjoint $\mathcal{E}^\dagger$ is unital, and in the last step we used \cref{eq:TraceD}. 

\sloppy A final fact one might note is that
\begin{align}
    \left| \mu(i,i'|\rho) \right| \in [0,1].
\end{align}
This is because from \cref{eq:muexplicit}, we have ${|\mu(i,i'|\rho)|=|\!\!\bra{a_i}\rho\ket{a'_{i'}}\!\!| |\!\!\braket{a'_{i'}}{a_i}\!\!|}$. The first term in the product satisfies ${|\!\!\bra{a_i}\rho\ket{a'_{i'}}\!\!|\leq 1}$, since the spectral decomposition ${\rho = \sum_k p_k \ketbra{\psi_k}{\psi_k}}$ gives $ {|\!\!\bra{a_i}\rho\ket{a'_{i'}}\!\!| =\sum_kp_k|\!\!\braket{a_i}{\psi_k}\!\!||\!\!\braket{\psi_k}{b_j}\!\!|\leq\sum_k p_k = 1}$. Then, since ${|\!\!\braket{a'_{i'}}{a_i}\!\!|\in[0,1]}$, we get ${|\mu(i,i'|\rho)|\in[0,1]}$. 

In fact, we can slightly generalize the results from Ref.~\cite{fernandes2024unitaryinvariant} from pure states to mixed states, to note that any complex number $\mu(i,i'|\rho) = |\Delta| e^{i\phi}$ must satisfy that 
\begin{equation}\label{eq:KD-phase_space_points}
    1-3|\Delta|^{\frac{2}{3}}+2|\Delta|\cos(\phi) \geq 0,
\end{equation}
for all possible $\vert a_i\rangle, \vert a'_{i'}\rangle$ and state $\rho$, with respect to any Hilbert space $\mathcal{H}_A$. This is true since $\mu(i,i'|\rho) = \text{Tr}[F_{i,i'}\rho]$ is linear with respect to $\rho$. Therefore, for any convex combination of pure states $\rho = \sum_\lambda \alpha_\lambda\ketbra{\psi_\lambda}{\psi_\lambda}$, we have that $\mu(i,i'|\rho)$ is \emph{also} a convex combination of pure state KD-phase space points  $\mu(i,i'|\psi_\lambda)$. Ref.~\cite{fernandes2024unitaryinvariant} showed that any $\mu(i,i'|\psi_\lambda)$ must be in the region described by Eq.~\eqref{eq:KD-phase_space_points}, but since this region forms a convex set of points in $\mathbb{C}$, any convex combination of terms $\mu(i,i'|\psi_\lambda)$ must also lie inside this region. This shows that the phases and absolute values of the numbers $\mu(i,i'|\psi_\lambda)$ satisfy non-trivial constraints---constraints which moreover relate to relevant physical facts. For example, maximal negativity is relevant for work extraction beyond classical thermodynamic limits~\cite{gherardini2024quasiprobabilities,hernandezgomez2023projective}. Eq.~\eqref{eq:KD-phase_space_points} implies that the maximal negative and imaginary values any KD distribution can attain are
$\min_{\{a_i\}_i,\{a'_{i'}\}_{i'},\rho} \text{Re}[\mu(i,i'|\rho)] = -1/8 $ and $\max_{\{a_i\}_i,\{a'_{i'}\}_{i'},\rho} \text{Im}[\mu(i,i'|\rho)] = 1/4$~\cite{fernandes2024unitaryinvariant}. 

Finally, we comment on our choice to only define KD representations using bases $\{\ket{a_i}\}_i$ and $\{\ket{a'_{i'}} \}_{i'}$ for the Hilbert space such that $\braket{a_i}{a'_{i'}}\neq 0$ for all $i$ and $i'$. This choice is fairly standard---but not universal---in the literature. Making this choice ensures that the representation we construct is faithful. As also pointed out in the main text, only in this case does learning a KD representation of a quantum process imply that one can reconstruct the given process, in which case ways of measuring the KD representation are simply ways of carrying out tomography (as recognized for the case of states in Ref.~\cite{johansen2007quantum}). Additionally, this property implies that the representation of the identity operator is the all-ones vector (Eq.~\eqref{unitrepn}), which is crucial for interpreting these KD distributions as quasistochastic in general---i.e., for normalization to be given by summing up all elements of the state.

\subsection{Relationship with frame representations and real-valued quasiprobability representations}

The terms {\em quasiprobability representation} and {\em frame-representation} sometimes refer only to real-valued representations of quantum theory, but are sometimes rather used to subsume complex-valued representations (such as KD representations). Similarly, the term frame is sometimes used specifically for bases for the space of Hermitian operators~\cite{Ferrie_2008,Ferrie_2009,posquasi,Schmid2024structuretheorem}; however, the origins of the term `frame' allow for arbitrary bases of more general inner product spaces~\cite[Sec. 1.1]{christensen2016introduction}, so one can use these terms in the context of complex-valued KD representations, as we have herein.

Frame representations are strictly more general than KD representations (as we have defined them).\footnote{Note however that some distributions under the KD moniker are not faithful (see e.g. Ref.~\cite[App. A]{arvidsson_Shukur2021conditions}), and so are not frame representations.} For one thing, KD representations use complete bases, while frame representations may use either complete or overcomplete frames (although the latter do not  represent the identity as the quasistochastic identity map~\cite{Schmid2024structuretheorem}). For another, the operator basis used in a standard KD representation is constructed to have a particular form (that in Eq.~\eqref{framedefn}), whereas {\em any} basis of operators may be used in a frame representation; for instance, the operators in the former case (but not in the latter) must be nonHermitian and rank one. 

\section{Proving stochasticity for all nonnegative KD representations}\label{stochproof}

We now prove that if a KD representation is real and nonnegative for some set of processes, then in fact it represents all those processes as substochastic matrices (and consequently, the representation constitutes a valid ontological model for the processes). 

We begin by showing that the representation of any state is a valid probability distribution in this case.
By assumption, $\mu(i,i'|\rho)\geq0$ for all $\rho$ under consideration, and by Eq.~\eqref{statenorm} we have that $\sum_{i,i'}\mu(i,i'|\rho) = 1$. So $\mu(i,i'|\rho)$ is a valid probability distribution.

\sloppy Next, we wish to show that a POVM element is in this case represented by a valid `response function': that is, a real  vector whose elements are between 0 and 1.
By assumption, $\xi(E|i,i')\geq0$ for all $E$ under consideration. Moreover, a POVM element $E$ can only arise in a theory or fragment of quantum theory as part of a measurement---a set of effects that sums to the identity effect $\mathds{1}$. Consequently, the POVM element $\mathds{1}-E$ is necessarily also in the scenario or fragment (one can measure this effect by doing any measurement containing POVM element $E$ and coarse-graining all the outcomes other than $E$). So (again by the assumption of nonnegativity) we have $\xi(E|i,i')\geq0$ and ${\xi(\mathds{1}-E|i,i')\geq0}$ for every $E$ under consideration. By linearity of the representation, the latter implies that $\xi(\mathds{1}|i,i')-\xi(E|i,i')\geq 0$. By \cref{unitrepn}, i.e., $\xi(\mathds{1}|i,i')=1$, we further have $1-\xi(E|i,i')\geq0$ and so
\begin{align}
 0 \leq \xi(E|i,i')\leq 1
\end{align}
for all $i,i'$ and for all $E$ under consideration, as required.

Finally, we show that under the assumption of nonnegativity, any KD representation of a quantum channel is a substochastic matrix. For quantum channels (which are trace-preserving), this is immediately evident from the assumption of nonnegativity together with Eq.~\eqref{tpnormcond}, namely
$\sum_{j,j'} \Gamma(j,j'|i,i',\mathcal{E}) 
= 1$ for all $i,i'$. If $\mathcal{E}$ is a trace-decreasing quantum operation, the argument is much the same as for effects: $\mathcal{E}$ can only arise in a scenario where it is complemented by some other complementary trace-decreasing quantum channel $\tilde{\mathcal{E}}$ such that $\mathcal{E}+\tilde{\mathcal{E}}$ is trace-preserving (so that each corresponds to the selective update rule for two possible outcomes of some quantum instrument). By linearity, one has 
\begin{align}
1 =&\sum_{j,j'}\Gamma(j,j'|i,i',\mathcal{E}+\tilde{\mathcal{E}}) \\ \nonumber
&= \sum_{j,j'}\Gamma(j,j'|i,i',\mathcal{E}) 
+\sum_{j,j'}\Gamma(j,j'|i,i',\tilde{\mathcal{E}})
\end{align}
for all $i,i'$. By assumption, all the terms in these sums are nonnegative, so none of them can be larger than 1. So for a trace-decreasing operation $\mathcal{E}$, one has
\begin{equation}
\sum_{j,j'}\Gamma(j,j'|i,i',\mathcal{E})\leq 1,
\end{equation}
as required for the representation to be substochastic.

\section{Proof of functoriality}\label{DPproof}

We now prove that every KD representation, as we have defined it, commutes with composition of quantum processes---formally, that it is a symmetric strict monoidal functor, or `is diagram-preserving'.

First, we prove that the quantum identity is represented as the identity on phase space:
\begin{align} 
&\Gamma(\bar{i},\bar{i}'|i,i',\mathcal{I}) = \Tr \left[F_{\bar{i},\bar{i'}}\mathcal{I}(D_{i,i'})\right] \nonumber\\ 
&= \Tr \left[F_{\bar{i},\bar{i}'}D_{i,i'}\right] = \delta_{\bar{i},i}\delta_{\bar{i}',i'},,
\end{align}
where we have used the duality of the frames, i.e., \cref{eq:DualityFD}.

Second, we prove that the KD representation preserves sequential composition, that is, for $\mathcal{E}_1$ transforming system A to system B and $\mathcal{E}_2$ transforming system B to system C, we have 

\begin{align}
    &\Gamma\bigl(k,k'|i,i',\mathcal{E}_2 \circ \mathcal{E}_1\bigr) \nonumber \\
    = &\sum_{j,j'} \Gamma(k,k'|j,j', \mathcal{E}_2) \Gamma(j,j'|i,i',\mathcal{E}_1).
\end{align}
Expanding the expression in the second line, we have
\begin{align}
    &\sum_{j,j'} \Gamma(k,k'|j,j', \mathcal{E}_2) \Gamma(j,j'|i,i', \mathcal{E}_1) \nonumber \\
 = & \sum_{j,j'} \Tr [F_{k,k'} {\cal E}_2( D_{j,j'})] \Tr [F_{j,j'} {\cal E}_1( D_{i,i'})] \nonumber \\
  =  & \Tr [F_{k,k'} {\cal E}_2 \left( \sum_{j,j'} \Tr [F_{j,j'} {\cal E}_1( D_{i,i'})] D_{j,j'}\right)]  \nonumber \\ 
   \stackrel{\eqref{eq_recOmu}}{=}&  \Tr [F_{k,k'} {\cal E}_2 \left( {\cal E}_1( D_{i,i'})\right)]  \nonumber \\
      = &  \Tr [F_{k,k'} ({\cal E}_2 \circ {\cal E}_1)( D_{i,i'})]  \nonumber \\
  = & \, \Gamma\bigl(k,k'|i,i',\mathcal{E}_2 \circ\mathcal{E}_1 \bigr),
\end{align}
where we used the linearity of the trace operation and the channel at the third equal sign, and the definitions of KD representations and the reconstruction equations at various places.

Third, we prove that KD representations preserve parallel composition, that is, for $\mathcal{E}_1$ transforming system A to system C and $\mathcal{E}_2$ transforming system B to system D, we have that the representation of the two channels in parallel, namely $\mathcal{E}_1\otimes\mathcal{E}_2$, satisfies
\begin{align}
    &\Gamma (k,k'|i,i',\mathcal{E}_1 ) \Gamma (l,l'|j,j',\mathcal{E}_2 ) \nonumber \\
    =& \Gamma (k,k';l,l'|i,i';j,j',\mathcal{E}_1\otimes\mathcal{E}_2 ). 
\end{align}
We expand the expression in the second line as
\begin{align}
    &\Gamma (k,k';l,l'|i,i';j,j',\mathcal{E}_1\otimes\mathcal{E}_2 ) \nonumber \\
    = & \Tr [F_{k,k';l,l'}{\cal E}_1\otimes {\cal E}_2 \left( D_{i,i';j,j'}\right)] \nonumber \\
  = & \Tr [\left(F_{k,k'} \otimes F_{l,l'}\right) {\cal E}_1\otimes {\cal E}_2 \left( D_{i,i'}\otimes D_{j,j'}\right)] \nonumber \\
    = & \Tr [F_{k,k'} {\cal E}_1( D_{i,i'}) \otimes  F_{l,l'} {\cal E}_1( D_{j,j'})] \nonumber \\
  = & \Tr [F_{k,k'} {\cal E}_1( D_{i,i'})]  \Tr [F_{l,l'} {\cal E}_1( D_{j,j'})] \nonumber \\
  = & \,\,\Gamma (k,k'|i,i',\mathcal{E}_1 )   \Gamma (l,l'|j,j',\mathcal{E}_2 ),
\end{align}
where the second equality follows from the definition of the frame and the dual frame for two systems (namely \cref{eq_def2frame,eq_def2dual}), and where the second to last equality uses a simple property of the trace, namely that $\Tr[A\otimes B]=\Tr[A]\Tr[B]$.

Last, we prove that the quantum swap channel is represented as the corresponding swap channel in the representation.
Consider the swap channel $\mathcal{S}$ such that
\begin{align}
    \mathcal{S} ( O_1 \otimes O_2 )= O_2 \otimes O_1,
\end{align}
where $O_1$ and $O_2$ are any operators, each of which is on a (potentially different) Hilbert space.  Note that the order $O_1\otimes O_2$ is reflected on $\Gamma$ via the order of the basis indices. We then need to show that in the KD representation of the swap, the order $i,i',j,j'$ of the input indices is swapped in the output to $\bar{j},\bar{j}',\bar{i},\bar{i}'$. Mathematically, 
\begin{align}
    & \Gamma (\bar{j},\bar{j}';\bar{i},\bar{i}'|i,i';j,j',\mathcal{S} )  \nonumber \\
    = & \delta_{\bar{i},i}\delta_{\bar{i}',i'}\delta_{\bar{j},j}\delta_{\bar{j}',j'},
\end{align}
for all indices. We expand the expression in the first line:
\begin{align}
    & \Gamma (\bar{j},\bar{j}';\bar{i},\bar{i}'|i,i';j,j',\mathcal{S} )   \nonumber \\
    = & \Tr [F_{\bar{j},\bar{j}';\bar{i},\bar{i}'} {\cal S}( D_{i,i';j,j'})] \nonumber \\
    = & \Tr \left[F_{\bar{j},\bar{j}'} \otimes F_{\bar{i},\bar{i}'}{\cal S} \left( D_{i,i'} \otimes D_{j,j'}  \right)\right] \nonumber \\
        = & \Tr \left[(F_{\bar{j},\bar{j}'} \otimes F_{\bar{i},\bar{i}'})(  D_{j,j'} \otimes D_{i,i'})\right] \nonumber \\
                = & \Tr \left[F_{\bar{j},\bar{j}'}D_{j,j'}\right] \Tr\left[  F_{\bar{i},\bar{i}'} D_{i,i'}\right] \nonumber \\
    = & \delta_{\bar{i},i}\delta_{\bar{i}',i'}\delta_{\bar{j},j}\delta_{\bar{j}',j'},
\end{align}
where the second equality follows from the definition of frame and dual frame for two systems (namely, \cref{eq_def2frame,eq_def2dual}).

These facts together imply \cite{mac2013categories} that every KD representation of quantum theory (as we have defined it) respects composition of quantum processes. Formally, such representations are symmetric strict monoidal functors---or in other words, diagram-preserving maps. 
    
\end{document}